\documentclass[11pt,titlepage]{article}
\usepackage[margin=1.2in]{geometry}
\usepackage{amsmath, amsbsy, amsthm, array, mathrsfs}
\usepackage{graphics}
\usepackage{physics}
\usepackage{epsfig, amssymb,latexsym,verbatim}
\usepackage{graphicx}
\usepackage{color}
\usepackage{dsfont, bbm}
\usepackage{relsize}
\usepackage{subcaption}

\newcommand{\R}{\mathbb{R}}
\newcommand{\mbs}{\mathbb{S}}

\newcommand{\noin}{\noindent}
\newcommand{\bee}{\begin{eqnarray*}}
\newcommand{\ene}{\end{eqnarray*}}
\newcommand{\bec}{\begin{center}}
\newcommand{\enc}{\end{center}}
\newcommand{\be}{\begin{equation}}
\newcommand{\ee}{\end{equation}}

\newcommand{\mc}{\mathcal}

\newcommand{\mb}{\mathbf}
\newcommand{\bs}{\boldsymbol}
\newcommand{\tb}{\textbf}
\newcommand{\pend}{$\blacksquare$}
\newcommand{\vs}{\vskip 3mm}
\newcommand{\bi}{\begin{itemize}}
\newcommand{\ei}{\end{itemize}}

\begin{document}

\title{\LARGE  Exact computation of projection regression depth and fast computation of its induced median and other estimators
 \\[4ex]
}

\author{ {\Large 
Yijun Zuo}\\[2ex]
         {\small {\em ~~ Department of Statics and Probability} }\\[2ex]
         {\small Michigan State University, East Lansing, MI 48824, USA} \\[2ex]
         {\small 
         zuo@msu.edu}\\[6ex]
     }
 \date{\today}
\maketitle

\vskip 3mm
{\small

\begin{abstract}
Zuo (2019) (Z19) addressed the computation of the projection regression depth (PRD)  and its induced median (the maximum depth estimator).
Z19 achieved the exact computation of PRD via a modified version of regular univariate sample median, which resulted in the
loss of invariance of PRD and the equivariance of depth induced median. This article achieves the exact computation without scarifying
the invariance of PRD and the equivariance of the regression median.
\vs
Z19 also addressed the approximate computation of  PRD induced median, the naive algorithm in Z19 is very slow.  This article modifies the approximation in Z19 and adopts Rcpp package and consequently obtains a much (could be $100$ times) faster algorithm with an even better level of accuracy meanwhile.\vs
 Furthermore, as the third major contribution, this article introduces three new depth induced
estimators which can run $300$ times faster than that of Z19 meanwhile maintaining the same 
 level of accuracy.
\vs

Real as well as simulated data examples are presented to illustrate the difference between the algorithms of Z19 and the ones proposed in this article. Findings support the statements above and manifest the major contributions of the article.

\bigskip
\par

\bigskip
\par
\noindent {\bf Running title:} Computation of projection regression depth and its induced estimators.
\end{abstract}
}
\section{Introduction}
Zuo (2019) (Z19) addressed the computation of projection regression depth (PRD) and its induced deepest estimator (aka regression median) $\bs{\beta}^*_{PRD}$ which were introduced in Zuo (2018) (Z18).
By modifying the definition of the univariate sample median, Z19 achieved the exact computation of the unfitness (UF) (defined is section 2), or equivalently the PRD. The approach, however, consequently scarifies the regression, scale, and affine invariance of PRD and the regression, scale, and affine equivariance of $\bs{\beta}^*_{PRD}$ (for related definitions, see Z18).\vs

A natural question is: can one compute the UF exactly without modifying the definition of univariate median and consequently keeping the very desirable properties? This article presents a positive answer to the question.\vs

Another major issue with Z19 is that the algorithm for computation of $\bs{\beta}^*_{PRD}$ is relatively very slow. 
Can the speed of the algorithm be improved so that it is more feasible and competitive in practice whereas the accuracy is maintained or even improved meanwhile?
\vs
 The second major contribution of this article is to introduce a much faster algorithm for $\bs{\beta}^*_{PRD}$, which can run in some cases more than $100$ times faster than that of Z19, meanwhile, always has a better accuracy or relative efficiency (i.e. smaller empirical mean squared error).\vs

The rest of the article is organized as follows.  Section 2 presents the projection regression depth (PRD) and its induced median $\bs{\beta}^*_{PRD}$. Section 3 addresses the exact computation of PRD  and approximate computation of $\bs{\beta}^*_{PRD}$.
Section 4 is devoted to the examples of the exact computation of PRD  as well as approximate computation of $\bs{\beta}^*_{PRD}$.
Section 5 introduces three PRD induced regression estimators that can run much (could be $300$ times) faster than that of Z19 meanwhile maintaining  small empirical mean squared errors, which constitutes the third major contribution of this article.\vs

Throughout, the linear regression model considered is:
\begin{eqnarray}
y&=&\mathbf{x}'\boldsymbol{\beta}+{{e}}, \label{eqn.model}
\end{eqnarray}
where  $'$ denotes the transpose of a vector, and random vector $\mathbf{x}=(x_1,\cdots, x_p)'$ and  parameter vector $\boldsymbol{\beta}$ are in $\R^p$ ($p\geq2$) and random variables $y$ and ${e}$ are in $\R^1$.
 If $\bs{\beta}=(\beta_0, \bs{\beta}'_1)'$ and $x_1=1$, then one has $y=\beta_0+\mb{x}'_1\bs{\beta}_1+{e}$, where $\mb{x}_1=(x_2,\cdots, x_p)' \in \R^{p-1}$.
 Let $\mb{w}=(1,\mb{x}'_1)'$. Then $y=\mb{w}'\bs{\beta}+{e}$. We use this model or (\ref{eqn.model}) interchangeably depending on the context.
\vs

\section{Projection regression depth and its induced median}
Z18 introduced  the PRD. For a given candidate  parameter $\bs{\beta}\in \R^p$, it is defined based on the UF (unfitness) as:
\be
\mbox{UF}(\bs{\beta};F_{(\mb{x}', y)})=\sup_{\mb{v}\in \mbs^{p-1}} \mbox{UF}_{\mb{v}}(\bs{\beta}; F_{(\mb{x}', y)}):=\sup_{\mb{v}\in \mbs^{p-1}}|{R}(F_{(\mb{w'}\mb{v},~ y-\mb{w'}\bs{\beta})})|\big/{S}(F_y),  \label{UF.eqn}
\ee
\vs
\be
\mbox{PRD}(\bs{\beta};F_{(\mb{x}', y)})=1/(1+\mbox{UF}(\bs{\beta};F_{(\mb{x}', y)})),\label{PRD.eqn}
\ee
\vs
\noin
where $F_{\mb{Z}}$ stands for the distribution of the d-dimensional random vector $\mb{Z} \in\R^d$ for any $d$,
$\mb{w'}=(1,\mb{x}')\in\R^p$, $\mbs^{p-1}=\{\mb{u}\in\R^p:~\|\mb{u}\|=1\}$. Throughout, $R$ will be restricted to the univariate regression functional  of the form $R(F_{(\mb{w'}\mb{v},~ y-\mb{w'}\bs{\beta})})=T\big(F_{(y-\mb{w}'\bs{\beta})/{\mb{w}'\mb{v}}}\big)$ and it is regression, scale, and affine equivariant (see page 116 of Rousseeuw and Leroy (1987)(RL87) for definitions). $T$ could be a univariate location functional that is location, scale and affine equivariant and $S$ is a scale functional that is translation invariant and scale equivariant (see pages 158-159 of RL87 for definitions), and $S(F_y)$ does not depend on $\mb{v}$ and $\bs{\beta}$.
\vs
   For robustness consideration, in the sequel, $(T,S)$ is the fixed pair $(\mbox{Med}, \mbox{MAD})$. That is the median (Med) and the median of absolute deviations (MAD) pair.
 Hereafter, we write $\text{Med}(Z)$ rather than
  $\text{Med}(F_Z)$.
For this special choice of $T$  and $S$  such that
\bee
 R(F_{(\mb{w'}\mb{v},~ y-\mb{w'}\bs{\beta})})&=&\text{Med}_{\mb{w'}\mb{v}\neq 0}\big(\frac{y-\mb{w}'\bs{\beta}}{\mb{w'}\mb{v}}\big),\\[1ex]\label{spesific-T.eqn}
 S(F_y)&=& \text{MAD}(F_y). \label{S.eqn}
 \ene
We have the \emph{unfitness} (UF) of $\bs{\beta}$ as
 \be
 \text{UF}(\bs{\beta}; F_{(\mb{x}', y)})=\sup_{\mb{v}\in \mbs^{p-1}}\Big|\text{Med}_{\mb{w'}\mb{v}\neq 0}\big(\frac{y-\mb{w}'\bs{\beta}}{\mb{w'}\mb{v}}\big)\Big|\bigg/ \text{MAD}(F_y),
 \ee
 and the \emph{projection regression depth} (PRD) of $\bs{\beta}$ as
 \be
 \text{PRD}\left(\bs{\beta}; F_{(\mb{x}', y)}\right)=\inf_{\mb{v}\in \mbs^{p-1},\mb{w'}\mb{v}\neq 0}
 \frac{\text{MAD}(F_y)}{\text{MAD}(F_y)+\Big|\text{Med}\big(\frac{y-\mb{w}'\bs{\beta}}{\mb{w'}\mb{v}}\big)\Big|}. \label{special-PRD.eqn}
 \ee
 \vs
 \noin
 Applying the min-max (or max-min) scheme, we obtain the maximum (deepest) \emph{projection regression depth estimating functional} (median) $\bs{\beta}^*_{PRD}$ (also denoted by $T^*_{\text{PRD}}$)
  w.r.t. the pair $(T,S)$
 \begin{eqnarray}
 \bs{\beta}^*_{PRD}(F_{(\mb{x}', y)})&=&\arg\!\min_{\boldsymbol{\beta}\in \R^p}\mbox{UF}(\boldsymbol{\beta}; ~F_{(\mb{x}', y)})  \label{eqn.T*}\\[1ex]
 &=&\arg\!\max_{\bs{\beta}\in\R^p}\text{PRD}\left(\boldsymbol{\beta}; ~F_{(\mb{x}', y)}\right).\nonumber
 \end{eqnarray}
PRD and $\bs{\beta}^*_{PRD}$ satisfy desirable properties, such as regression, scale and affine invariance and equivariance, respectively, see Z18
for definitions and more detailed discussions. These desirable properties will be deprived  in the empirical case when the sample median is modified as did in Z19.
\section{Computational problems}
\subsection{Exact computation of PRD} \label{EX-prd.section}
For a given $\bs{\beta}$ and sample $\mb{Z}^{(n)}=\{(\mb{x}'_i, y_i), i=1,\cdots, n\}$ in $\R^p$, the computation of PRD$(\bs{\beta}, F^n_{\mb {Z}})$, or equivalently  of UF$(\bs{\beta}, F^n_{\mb{Z}})$,  is to compute the quantity below:
\be
\mbox{UF}(\bs{\beta}; F^n_{\mb{Z}})=\sup_{\mb{v}\in \mbs^{p-1}}\Big|\text{Med}_{\mb{w}'_i\mb{v}\neq 0}\big\{\frac{y_i-\mb{w}'_i\bs{\beta}}{\mb{w}'_i\mb{v}}\big\}\Big|\bigg/S_y, \label{UF-1.eqn}
\ee
where $F^n_{\mb{Z}}$ is the empirical distribution based on $\mb{Z}^{(n)}$,  $\mb{w}'_i=(1,\mb{x}'_i)$ and $S_y=\text{MAD}_i\{y_i\}$. Hereafter we assume that \tb{(A1)}: $P(\mb{w}'\mb{v}=0)=0$,~$\forall~ \mb{v}\in\mbs^{p-1}$; 
 and \tb{(A2)} $P(r(\bs{\beta})=0)=0$, where $r(\bs{\beta})=y-\mb{w}'\bs{\beta}$,
~$\forall ~\bs{\beta}\in \R^p$.
\tb{(A1)-(A2)} hold automatically if $(\mb{x}', y)'$ has a density, or if $\mb{x}$ does not concentrate on a single $(p-2)$ dimensional hyperplane in $\mb{x}$ space and any $(p-1)$ dimensional hyperplane determined by $r(\bs{\beta})=0$ in $(\mb{x}', y)'$ space does not contain any probability mass.
\vs

For the simplicity of description, we write $\mb{t}'_i=\mb{w}_i'/r_i(\bs{\beta})$,  where 
$r_i(\bs{\beta})=y_i-\mb{w}'_i\bs{\beta}$. Now the computation of $\mbox{UF}(\bs{\beta}; F^n_{\mb{Z}})$ in (\ref{UF-1.eqn}) is equivalent to
the computation of
\be
\mbox{UF}(\bs{\beta}; F^n_{\mb{Z}})=\sup_{\mb{v}\in \mbs^{p-1}}\bigg|\text{Med}_{\mb{t}'_i\mb{v}\neq 0 }\big\{\frac{1}{\mb{t}'_i\mb{v}}\big\}\bigg|\bigg/S_y .\label{UF-2.eqn}
\ee

Again for the simplicity of description, we write $k^{\mb{v}}_i=1/\mb{t}_i'\mb{v}$ and $u^{\mb{v}}_i=\mb{t}_i'\mb{v}$. The latter two are well defined almost surely (a.s.) under \tb{(A1)-(A2)}.  Without loss of generality, 
hereafter assumes  that $S_y=1$ (since it does not depend on $\mb{v}$ or $\bs{\beta}$). The UF$(\bs{\beta};F^n_{\mb{Z}})$ in (\ref{UF-2.eqn}) is then 
\be
\mbox{UF}(\bs{\beta}; F^n_{\mb{Z}})=\sup_{\mb{v}\in\mbs^{p-1}}\bigg|\text{Med}_i\big\{k^{\mb{v}}_i\big\} \bigg| :=\sup_{\mb{v}\in\mbs^{p-1}}\bigg|g(\mb{v}) \bigg|.
\label{UF-3.eqn}
\ee
\vs
The exact computation of (\ref{UF-3.eqn}) above is still very challenging if it is not impossible.
Let $k^{\mb{v}}_{(1)}\leq k^{\mb{v}}_{(2)}\cdots\leq k^{\mb{v}}_{(n)}$ be ordered values of $k^{\mb{v}}_i$.
 Partition $\mbs^{p-1}$ into two disjoint parts
\be
\mc{S}_1=\{\mb{v}\in \mbs^{p-1}:~ k^{\mb{v}}_{(1)}<0 ~\mbox{and}~ k^{\mb{v}}_{(n)}>0\};~~~\mc{S}_2=\{\mb{v}\in\mbs^{p-1}:~ k^{\mb{v}}_{(1)}> 0 ~\mbox{or}~ k^{\mb{v}}_{(n)}< 0\}. \label{Si.eqn}
\ee
It is readily seen that both $\mc{S}_1$ and $\sc{S}_2$ are symmetric about the origin. That is, if $\mb{v} \in \mc{S}_i$ then, $\mb{-v} \in \mc{S}_i$.
Now the UF$(\bs{\beta};F^n_{\mb{Z}})$ in (\ref{UF-3.eqn})  can be expressed as follows:
\be
\mbox{UF}(\bs{\beta}; F^n_{\mb{Z}})=\max\big\{\sup_{\mb{v}\in \mc{S}_1}|g(\mb{v})|, ~\sup_{\mb{v}\in \mc{S}_2}|g(\mb{v})|\big\}. \label{UF-4.eqn}
\ee
\vs
 For a given sample $\mb{Z}^{(n)}:=\{(\mb{x}'_i, y_i), ~ i=1,\cdots, n\}$,  $\bs{\beta}$ in $\R^p$ and $\mb{v}\in\mbs^{p-1}$, since $k^{\mb{v}}_{(1)}\leq k^{\mb{v}}_{(2)}\leq \cdots \leq k^{\mb{v}}_{(n)}$ are ordered value of $k^{\mb{v}}_{i}= 1/\mb{t}'_i\mb{v}$, then $1/\mb{t}'_{i_1}\mb{v}\leq 1/\mb{t}'_{i_2}\mb{v}\leq \cdots \leq 1/\mb{t}'_{i_n}\mb{v}$ for some $\{i_1,\cdots, i_n\}$, a permutation of $\{1, 2, \cdots, n\}$.
 Similarly, $u^{\mb{v}}_{(1)}\leq u^{\mb{v}}_{(2)}\leq \cdots \leq u^{\mb{v}}_{(n)}$  corresponds to a permutation $\{j_i,\cdots, j_n\}$ such that
  $u^{\mb{v}}_{j_1}\leq u^{\mb{v}}_{j_2}\leq \cdots,\leq u^{\mb{v}}_{j_n}$ for  $u^{\mb{v}}_i=\mb{t}'_i\mb{v}$.

 \vs
\noin
\tb{Proposition 3.1}: Assume \tb{(A1)-(A2)} hold. Let $N^{-}_{\mb{v}}:=\sum_{i=1}^n\mb{I}(k_i^{\mb{v}}<0)$.  The unfitness of $\bs{\beta}$ in (\ref{UF-1.eqn}) can be computed equivalently via 
(\ref{UF-4.eqn}). The latters can be computed
as follows.
\vs
\noin
Denote $n1:=\lfloor(n+1)/2\rfloor$ and $n2:=\lfloor(n+2)/2\rfloor$, where $\lfloor \cdot\rfloor$ is the floor function.  \vs
\noin
(i) For $\mb{v} \in \mc{S}_2$, 
\[
\sup_{\mb{v}\in \mc{S}_2}|g(\mb{v})|= \left\{
\begin{array}{ll}
\max_{\mb{v}\in \mc{S}_2} \frac{(\mb{t}'_{i_{n1}}+\mb{t}'_{i_{n2}})\mb{v}\big/2} {\mb{v}'\mb{t}_{i_{n1}}\mb{t}'_{i_{n2}}\mb{v}}
 &\mbox{~if~} N^{-}_{\mb{v}}=0, 
\\[3ex]
-\min_{\mb{v}\in \mc{S}_2} \frac{(\mb{t}'_{i_{n1}}+\mb{t}'_{i_{n2}})\mb{v}\big/2} {\mb{v}'\mb{t}_{i_{n1}}\mb{t}'_{i_{n2}}\mb{v}} &\mbox{~if~} N^{-}_{\mb{v}}=n. 
\end{array}
\right.
\label{UF-34.eqn}
\]
\vs
\noin
(ii) For $\mb{v}\in \mc{S}_1$, let $m$ be a non-negative integer.
\begin{itemize}
\item[]
 if $n=2m+1$,
\[
\sup_{\mb{v}\in \mc{S}_1}|g(\mb{v})|=\left\{
\begin{array}{ll}
-1\Big/\max_{\mb{v}\in \mc{S}_1} \mb{t}'_{i_{n1}}\mb{v}      
& \mbox{if $k^{\mb{v}}_{(n1)}<0$},\\[2ex]
1\Big/\min_{\mb{v}\in \mc{S}_1}\mb{t}'_{i_{n1}}\mb{v}                     
& \mbox{if $k^{\mb{v}}_{(n1)}>0$},
\end{array}
\right.
\]
\item[] if $n=2m+2$,
\[
\sup_{\mb{v}\in S_1}|g(\mb{v})|=\left\{
\begin{array}{ll}
\bigg|\max_{\mb{v}\in \mc{S}_1} \frac{(\mb{t}'_{i_{n1}}+\mb{t}'_{i_{n2}})\mb{v}\big/2}{ \mb{v}'\mb{t}_{i_{n1}}\mb{t}'_{i_{n2}}\mb{v} }  \bigg| &
~\mbox{if}~~ k^{\mb{v}}_{(n1)}<0 ~\mbox{and}~ k^{\mb{v}}_{(n2)}>0, \\[2.5ex]
\max_{\mb{v} \in \mc{S}_1} \frac{\big(\mb{t}'_{i_{n1}}+\mb{t}'_{i_{n2}}\big)\mb{v}\big/2}
{\mb{v}'\mb{t}_{i_{n1}}\mb{t}'_{i_{n2}} \mb{v}}
 & ~\mbox{if}~~k^{\mb{v}}_{(n1)}>0,\\[3.5ex]
-\min_{\mb{v}\in \mc{S}_1}
\frac{\big(\mb{t}'_{i_{n1}}+\mb{t}'_{i_{n2}}\big)\mb{v}\big/2}
{\mb{v}'\mb{t}_{i_{n1}}\mb{t}'_{i_{n2}} \mb{v}}
& ~\mbox{if}~~ k^{\mb{v}}_{(n2)}<0.
\end{array}
\right.
\]
\end{itemize}
\vs

\noindent
\tb{Proof:}  Note that  under \tb{(A1)-(A2)} $\mc{S}_i$ ($i=1,2$) are closed sets a.s..  In light of Proposition 2.1 and Corollary 2.1 of Z19 and the proofs there, the proof here follows immediately.   Details are straightforward to verify and thus are omitted. \hfill \pend
\vs
\noin
\tb{Remarks 3.1}:
The proposition gives a clear foundation for the exact computation of UF$(\bs{\beta}, F^n_{\mb{Z}})$, or equivalently PRD($(\bs{\beta}, F^n_{\mb{Z}})=\big(1+\mbox{UF}(\bs{\beta}, F^n_{\mb{Z}})\big)^{-1}$.
\bi
\item[(I)]  UF$(\bs{\beta}, F^n_{\mb{Z}})$ can be exactly computed via the optimization over closed sets $\mc{S}_i$. There are unified formulas over $\mc{S}_i$ for distinct cases of permutations. And two types of optimization problems exist in the proposition
\bi
\item[(i)]\tb{Type I}: $\min$ (or $\max$) of $\mb{c}'\mb{v}$ for $\mb{v} $ over a closed subset set of $\mbs^{p-1}$ and $\mb{c}\in\R^p$.
\item[(ii)] \tb{Type II}: $\min$ (or $\max$) of $\frac{\mb{b}'\mb{v}}{\mb{v}'\mb{A}\mb{v}}$ for $\mb{v}$ over a closed subset set of $\mbs^{p-1}$ and $\mb{b}\in\R^p$, $\mb{A}\in \R^{p\times p}$ ($A$ is symmetric and positive-definite over the set).
\ei
\item[(II)]
 $\mb{b}$, $\mb{c}$, and $\mb{A}$ above are determined by $\{\mb{t}_i\}$ and depend on $\mb{v}$ only through the permutation $i_1,\cdots, i_n$ which is induced by the projection of  $\{\mb{t}_i\}$ onto $\mb{v}$. That is, for a given sample and $\bs{\beta}\in \R^p$, and a $\mb{v}\in \mbs^{p-1}$ or more generally a fixed permutation $i_1,\cdots, i_n$ (of $\{1, 2,\cdots, n\}$) over a set of $\mb{v}$, $\mb{b}$, $\mb{c}$, and $\mb{A}$ are constant vectors and matrix.\vs

 Hence, with the constraints discussed in the sequel,  \tb{Type I} optimization could be solved by linear programming and \tb{Type II} optimization could be solved by gradient-type, Newton-type, or interior-point methods (see, e.g. Numerical Recipes (2007) Chapter 10, Freund (2004), and Boyd and Vandenberghe (2004)), among others.
 \item[(III)] When $n$ is odd, there is just one type, \tb{Type I}, optimization problem. The exact computation is much easier. To deal with even $n$ case, Z19 modified the definition of the regular sample median (adopted the lower median) to simplify the exact computation to just a \tb{Type I} optimization problem.
 \hfill \pend
\ei
\noin
To get the exact value of UF$(\bs{\beta}, F^n_{\mb{Z}})$ utilizing the proposition, 
 it seems that one has to know the set $\mc{S}_i$ first, $i=1,2$ (or more accurate their boundaries).   $\mc{S}_2$ can be empty. In fact, when the convex hull formed by all $\mb{t}_i$'s contains the origin, then $\mc{S}_1=\mbs^{p-1}$. Fortunately, we do not have to identify $\mc{S}_i$, $i=1,2$.  \vs

Since there is no unique formula over $\mc{S}_i$ in  proposition, therefore, for the exact computation task, we have to further partition $\mc{S}_i$ into disjoint pieces. For example, partition $\mc{S}_1$ into five pieces and $\mc{S}_2$ into two pieces, according to the cases listed in the proposition 3.1.  The latter task is not easier than identifying $\mc{S}_i$. 
    For example, identify all $\mb{v} \in\mc{S}_1$ such that $k^{\mb{v}}_{(n1)}>0$ for even $n$ case is not straightforward at all. we seek other approaches below.\vs

For a given sample $\mb{Z}^{(n)}$ and $\bs{\beta} \in \R^p$ and $\mb{v} \in \mbs^{p-1}$, there is a unique permutation $i_1,\cdots, i_n$ of $\{1,2,\cdots, n\}$ induced by $k^{\mb{v}}_i=1/ \mb{t}'_i \mb{v}$. The  $k^{\mb{v}}_{i_j}$ ($j=1,\cdots, n$) are all we need for the calculation in (\ref{UF-3.eqn}) or Proposition 3.1.  However, a permutation $i_1,\cdots, i_n$ corresponds to a set of $\mb{v}\in\mbs^{p-1}$ each of them can produce the same permutation via $\{k^{\mb{v}}_i\}$.
\vs
That is, a fixed permutation corresponds to a unique piece of $\mbs^{p-1}$ (or of the surface of the unit sphere). There are
totally at most  $n!$ possible permutations hence $n!$ disjoint pieces that partition the $\mbs^{p-1}$ (or the surface of the unit sphere).
By proposition 2.2 of Z19, each piece  belongs to either $\mc{S}_1$ or $\mc{S}_2$.
Selecting one $\mb{v}$ from each piece suffices for the exact computation of UF$(\bs{\beta}, F^n_Z)$ via Proposition 3.1.
The cost is approximate of order $O(n^{n+1/2})$ without counting optimization cost, unaffordable magnitude of cost. We seek to merge some pieces. 
\vs
 In light of the Observation 3 (O3) in Z19 on $\mb{v}$ induced permutations (the circular or spherical sequence), when $\mb{v}$ moves on the surface of the unit sphere, its induced permutation changes only when it crosses a hyperplane ($H_0$) that goes through the origin and is perpendicular to another hyperplane ($H_1$) that is formed by 
 sample points from $\{\mb{t}_i\}$.
 \vs
  The former hyperplanes ($H_0$'s) (each contains the origin) cut  the $\mbs^{p-1}$ into disjoint $N(n,p)$ pieces $P_k$  ($k=1,\cdots, N(n, p)$), where $N(n, p):= 2\sum_{i=0}^{p-1}{q-1\choose i}$ (see Winder(1966)) and $q:=N_n^p(\{\mb{t}_i\})$ is the total distinct $(p-1)$-dimensional hyperplanes formed by 
  points from $\{\mb{t}_i\}$. $q \leq {n \choose p}$. Assume $q> 1$. When $\mb{t}_1,\cdots, \mb{t}_n$ are in a general position (see Z19 for definition), $q={n \choose p}$. In the latter case, $N(n, p)=O(n^{p(p-1)})$, smaller than $O(n^{n+1/2})$ above if $n\geq p (p-1)$.
  \vs

  Each $P_k$ ($k=1,\cdots, N(n,p)$) corresponds to a unique permutation $\{i_1,\cdots, i_n\}$, that is, $1/\mb{t}'_{i_1}{\mb{v}_0}\leq 1/\mb{t}'_{i_2}{\mb{v}_0}\leq \cdots \leq 1/\mb{t}'_{i_n}{\mb{v}_0}$, $\forall~ \mb{v}_0\in P_k$. The latter in turn corresponds to  a polyhedral cone (see Z19) which is determined by
 \be
   \mb{B}'\mb{v}\leq \mb{0}_{(n-1)\times 1}, \label{B.eqn}
 \ee
 where $\mb{v}\in \mbs^{p-1}$ and $B=(B_1,\cdots, B_{n-1})_{p \times (n-1)}$, $B_j:= \mb{t}_{i_j}-\mb{t}_{i_{j+1}}$, $j=1,\cdots, N^{-}_{\mb{v}_0}$; $B_j:= -(\mb{t}_{i_j}-\mb{t}_{i_{j+1}})$, $j=N^{-}_{\mb{v}_0}+1,\cdots, (n-1)$, and vector inequality is in the coordinate-wise sense.
\vs
By proposition 2.2 of Z19, the entire $P_k$ belongs to only one of $\mc{S}_i$. So as long as we have one $\mb{v}_0$ from each $P_k$, we can easily produce the permutation associated with $P_k$ and the induced $k^{\mb{v_0}}_i$ and determine which $\mc{S}_i$ and formulae should use in Proposition 3.1. Coupled with the constraints $B'\mb{v}\leq \mb{0}_{(n-1)\times 1}$ above, both Type I and Type II optimization problems in the proposition could be solved in linear time (note that $\mb{b},\mb{c}$ and $\mb{A}$ are constants over the entire piece of $P_k$). The exact computation of UF$(\bs{\beta},F^{n}_{\mb{Z}})$ could be achieved with the worst-case time complexity of order $TC(n,p,N_{iter}):=O( N(n,p)(p^{2.5}+n\log n+np^{1.5}+npN_{iter}))$, where $N_{iter}$ is the number of iterations needed when solving the type II optimization problem.
\vs
\noin
\tb{Theorem 3.1} Under \tb{(A1)-(A2)}, for a given sample $\mb{Z}^{(n)}$ and a $\bs{\beta}$ in $\R^p$, UF$(\bs{\beta},F^n_{\mb{Z}})$ (or PRD$(\bs{\beta}, F^n_{\mb{Z}})$)
can be computed exactly with the worst-case complexity of $TC(n,p,N_{iter})$.\vs
\noin
\tb{Proof}: ~
Obviously, exact computation is achieved if we can obtain the RHS of display (\ref{UF-4.eqn}). For the latter, we appeal to Proposition 3.1.
To implement the proposition, essentially,  we need to solve the two types of optimization problems in the proposition.\vs
By the discussion immediately before the theorem, we know the key for the optimization problems is to identify all pieces  $P_k$ ($k=1,\cdots, N(n,p)$) of $\mbs^{p-1}$. Equivalently, to identify all $N(n,p)$ distinct permutations of $\{1,2,\cdots, n\}$. The latter is equivalently to find a unit vector $\mb{u}\in P_k$ for each $P_k$ which can produce the unique fixed permutation over $P_k$.
\vs
Each $P_k$ is the intersection of $\mbs^{p-1}$ and the
polyhedron cone formed by the constraint $B'\mb{v}\leq \mb{0}_{(n-1)\times 1}$. The edge (or ridge) of the cone can be used to find the $\mb{u}$ above, which is shared by another adjacent cone. In other words, it is the intersection of (at least) two hyperplanes $H_0$'s which go through the origin and are perpendicular to two hyperplanes $H_1$'s each of which is formed by  points from $\{\mb{t}_i\}$, respectively.
\vs
The direction from the origin to any other point on the intersection hyperline of two hyperplanes $H_0$'s is the solution of the vector sought.
 Denote the direction by $\mb{u}$ ($\mb{u}$ could also be obtained more costly via the origin and any vertex of the cone through vertex enumeration
 (see Bremner et al., 1998, Paindaveine and \' Siman (2012) and Liu and Zuo (2014))).\vs
Each $\mb{u}$ above lies on the boundary of $P_k$. It not only lies in the facet of one cone but also
 lies in that of an adjacent cone which shares the common intersection hyperline (edge or ridge) with the former cone. Tiny perturbation
 of $\mb{u}$ in opposite directions will lead $\mb{u}$ entering the interiors of the two adjacent cones. There might be more than two cones that are adjacent. Thus, every $\mb{u}$ might yield two or more new permutations (the scheme in the algorithm yields up to $8 \times(p-2)$ distinct ones, $p> 2$). \vs


 Update the total number $N_{permu}$ of distinct permutations. 
With respect to each distinct permutation, or equivalent over each $P_k$, update $\sup_{v\in\mbs^{p-1}}|g(\mb{v})|$ according to Proposition 3.1 and carry out one of the two types of optimization.\vs
Repeat above steps until $N_{permu}=N(n,p)$ or UF could not be improved after trying $\kappa p$ more distinct permutations ($\kappa$ is a positive integer, could be, say, $10, 20$, or even $50$)
\vs

The cost of computation of each of elements of the descriptions above is as follows.
\bi
\item[(a)] obtaining all $\{\mb{t}_i\}$ costs $O(np)$,
\item[(b)] calculating normal vectors $\mb{v}_i$ of $H^i_1$ and normal vectors $\mb{u}_i$ of $H^i_0$ ($i=1, 2$), and $\mb{u}$ that is perpendicular to $\mb{u}_i$, the total cost is $O(p^3)$,
\item[(c)] producing each permutation costs $O(n(p+\log n))$,
\item[(d)] updating $\sup_{v\in\mbs^{p-1}}|g(\mb{v}|$ according to Proposition 3.1 costs $O(n\log n)$,
\item[(e)] linear programming is $O(p^{1.5}n+p^{2.5})$ (see Yin Tat Lee and Aaron Sidford (2015) which is even further improved by Cohen, Lee, and Song (2019)),
\item[(f)] for the type II non-convex and nonlinear optimization problem, one can use the conjugate gradient method or even better the primal-dual interior-point method (Wright (1997), Morales, et al (2003)) combined with the sequential quadratic programming (SQP) ( Nocedal and Wright (2006)), e.g. package LOQO (Vanderbei and Shanno (1999) and  Vanderbei (1999)) with cost
$O(npN_{iter})$, where $N_{iter}$ is the number of iterations needed in LOQO.
\ei
\vs
Keeping only the dominating terms, we thus have the overall worst-case time complexity $TC(n,p, N_{iter})=O(N(n,p)(n\log n+np^{1.5}+p^{2.5}+npN_{iter}))$.
\hfill \pend
\vs
\noin
\tb{Pseudocode} (Exact computation of UF$(\bs{\beta},F^n_{\mb{Z}})$, or equivalently of PRD$(\bs{\beta}, F^n_{\mb{Z}})$)
\bi
\item Calculate $\{\mb{t}_i\}$ and $N(n,p)$ (assume that $\{\mb{t}_i\}$ are in general position); set $N_{permu}
=\mbox{UF}=0$.
\item While ($N_{permu}<N(n,p)$)
\bi
\item[1] Obtain $\mb{u}$ and its induced permutations, store (and update $N_{permu}$ of the total number of) the distinct permutations.
\item[2] Update UF$=\sup_{\mb{v}\in\mbs^{p-1}}|g(\mb{v})|$ via proposition 3.1 and carry out the corresponding optimization.
\item[3] 
 If UF could not be improved after trying $\kappa p$ more distinct permutations, break the loop.
\ei
\item Output UF (or $1/(1+\mbox{UF})$). \hfill \pend
\ei
\vs
\noin
\tb{Remarks 3.2}\vs
\tb{(I)} In the best scenario, $N(n,p)$ could be replaced by  $O(n^2)$ (
if $p=2$). Even in this case, the cost of exact computation is $O(n^2(n\log n+np^{1.5}+p^{2.5}+npN_{iter}))$, which is still unaffordable for large $n$ and $p$. An approximate algorithm, such as \tb{AA-UF-3} of Z19 with cost of order $O(N(np+p^3) + np)$, where tuning parameter $N$ being the total number of normal directions of the hyperplanes formed by p points from $\{\mb{t}_i\}$, is more feasible in practice. \vs
\tb{(II)} By altering the definition of the traditional sample median (using the ``low median"),  Z19 also achieved the exact computation of UF$(\bs{\beta},F^n_{\mb{Z}})$ and proposed an algorithm that has slightly less cost (no term of $npN_{iter}$). For this  advantage, it pays a price of losing affine invariance of resulting PRD and the affine equivariance of induced $\bs{\beta}^*_{PRD}$, nevertheless. \vs Furthermore, $\bs{\beta}^*_{PRD}$ in Z19 can no longer recover the traditional univariate median when $p=1$. That is, the
maximum regression depth estimator in Z19 is not a generalization of univariate median to regression in a multi-dimensional setting. We show that in next section,
our current version of $\bs{\beta}^*_{PRD}$ does recover the univariate median when $p=1$.
\hfill  \pend
\subsection{Approximate computation of PRD induced median}
Before addressing the approximate computation of maximum projection depth estimator (or median), we first show that it indeed deserves to be called a median since it recovers the univariate sample median when $p=1$.
Recall that (assume, without the loss of generality (w.l.o.g), again $S_y=1$)
\be
\bs{\beta}^*_{PRD}=\arg\min_{\bs{\beta}\in \R^p}\sup_{\mb{v}\in\mbs^{p-1}}\Big|\mbox{Med}_i\{\frac{y_i-\mb{w}'_i\bs{\beta}}{\mb{w}'_i\mb{v}}\} \Big|.
\label{beta*.eqn}
\ee
When $p=1$, it reduces to the following
\be
{\beta}^*_{PRD}=\arg\min_{{\beta}\in \R}\sup_{{v}=\pm 1}\Big|\mbox{Med}_i\{\frac{y_i-{\beta}}{{v}}\} \Big|. \label{beta-cover-median.eqn}
\ee
 We have\vs
\noin
\tb{Proposition 3.2} 
When $p=1$, 
the ${\beta}^*_{PRD}$ recovers to the regular sample median of $\{y_i\}$.\vs
\noin
\tb{Proof}:
~
Let $y_{(1)}\leq y_{(2)}\leq \cdots \leq y_{(n)}$ be the ordered values of $y_i$ and $\mu=(y_{(n1)}+y_{(n2)})/2$, where $n1$ and $n2$ are defined in proposition 3.1, i.e., $\mu$ is the regular sample median of $\{y_i\}$. We show that $\beta^*_{PRD}=\mu$. 
It is readily seen that 
\be
{\beta}^*_{PRD}=\arg\min_{{\beta}\in \R}\big|\mbox{Med}_i\{{y_i-{\beta}}\}\big|=
\arg\min_{{\beta}\in \R}\big|\mbox{Med}_i\{y_i\}-{\beta}\big|= \arg\min_{{\beta}\in \R}\big|\mu-{\beta}\big|,
 \label{proof-cover-medain.eqn}
\ee
where the first equality follows from (\ref{beta-cover-median.eqn}) and the oddness of median operator, the second one follows from the translation equivalence (see page 249 of RL87 for definition) of the median as a location estimator,
the third one follows from the definition of $\mu$.\vs
The RHS of (\ref{proof-cover-medain.eqn}) above indicates that $\mu$ is the only solution for $\bs{\beta}^*_{PRD}$.
\hfill \pend
\vs
\noin
\tb{Remarks 3.3}\vs
\tb{(I)} The proposition holds true for the univariate population median. That is, $\bs{\beta}^*_{PRD}$ also recovers the univariate median in the population case.\vs
\tb{(II)} If one modifies the definition of Med in the UF as did in Z19, then the proposition no longer holds true in both sample or population cases.
\hfill \pend
\vs \vs

Now we turn to the approximate computation of $\bs{\beta}^*_{PRD}$ in (\ref{beta*.eqn}). First, we notice that the $\bs{\beta}^*_{PRD}$ must be bounded, or equivalently, the search for the optimal $\bs{\beta}$ in the RHS of (\ref{beta*.eqn}) could be limited within a bounded set (hypersphere). To see this, notice that for a given $\bs{\beta}\neq 0$, let $\mb{v}_0=\bs{\beta}/\|\bs{\beta}\|$, then
\bee
\mbox{UF}(\bs{\beta}, F^n_Z)&=& \sup_{\mb{v} \in\mbs^{p-1}}\Big| \mbox{Med}_i\{(y_i-\mb{w}_i'\bs{\beta})/\mb{w}'_i\mb{v}\}\Big| 
\geq \Big| \mbox{Med}_i\{(y_i-\mb{w}_i'\bs{\beta})/\mb{w}'_i\mb{v}_0\}\Big|\\[1ex]
&=& \Big| \mbox{Med}_i\{y_i/\mb{w}'_i\mb{v}_0\}- \|\bs{\beta}\|\Big| 
\longrightarrow \infty ~(a.s.), \mbox{~~~~as $\|\bs{\beta}\|\to \infty$,}
\ene
where the last step follows from the fact that $\mbox{Med}_i\{y_i/\mb{w}'_i\bs{\beta}\}=1$ cannot hold for any $\bs{\beta}$ with $\|\bs{\beta}\|\to \infty$. Let $\delta=\mbox{UF}(\mb{0},F^n_Z)$ and $c^*=\sup\{\|\bs{\beta}\|, \mbox{UF}(\bs{\beta}, F^n_Z)\leq \delta\}$,  
then we have
$$
\bs{\beta}^*_{PRD}=\arg\min_{\|\bs{\beta}\|\leq c^*}\sup_{\mb{v}\in\mbs^{p-1}}\Big|\mbox{Med}_i\Big\{\frac{y_i-\mb{w}'_i\bs{\beta}}{\mb{w}'_i\mb{v}}\Big\}\Big|.$$

In light of RHS of the display above,  the generic steps of compute $\bs{\beta}^*_{PRD}$ are listed as follows:
\bi
\item[(A)] Select randomly $p$ points from $\mb{Z}^{(n)}=\{(\mb{x}'_i, y_i)\} \in \R^p$, which determine a $\bs{\beta}$ through the hyperplane $y=\mb{w}'\bs{\beta}$. Produce a set of $N_{\bs{\beta}}$ possible $\bs{\beta}$'s: $S_{\bs{\beta}}=\{\bs{\beta}_1,\cdots, \bs{\beta}_{N_{\bs{\beta}}}\}$  in this way, where $N_{\bs{\beta}}$
    is a tuning parameter (could be, e.g., $N_{\bs{\beta}}=\min\{1000, {n \choose p}\}$).
\item[(B)] Let $S^1_{\mb{v}}=\{\mb{v}_i \in \mbs^{p-1}, i=1,\cdots, N_{\mb{v}}\}$, where $\mb{v}_i$ are normal vectors to the hyperplanes in (A).
Let $S^2_{\mb{v}}=\{\mb{v}_i \in \mbs^{p-1}, i=1,\cdots, N_{\mb{v}}\}$, where $\mb{v}_i$ is the normal vector to the hyperplane formed by $p$ points from $\{\mb{w}_i/r_i(\bs{\beta})\}$, where $N_{\mb{v}}$ is another tuning parameter.
 Let $S^3_{\mb{v}}=\{\mb{v}^j_i:=\frac{\bs{\beta}_j-\bs{\beta}_i}{\|\bs{\beta}_j-\bs{\beta}_i\|}, ~\forall ~ \bs{\beta}_i(\neq \bs{\beta}_j)\in S_{\bs{\beta}}\}$ for some $\bs{\beta}_j\in S_{\bs{\beta}}$. Set $S_{\mb{v}}=S^1_{\mb{v}}\cup S^2_{\mb{v}} $ plus some $\mb{v}$'s from $S^3_{\mb{v}}$.
\item[(C)] 
Search over
the convex hull formed by all $\bs{\beta}\in S_{\bs{\beta}}$ (or the ball $\|\bs{\beta}\|\leq c^*$) for the $\bs{\beta}$ that has the minimum approximate unfitness (using all $\mb{v} \in S_{\mb{v}}$ for the calculation for the unfitness). The $\bs{\beta}$  serves as an approximate $\bs{\beta}^*_{PRD}$.
\item[(D)] To mitigate the effect of randomness, repeat steps (A)-(C) above many times to get the final overall best approximate $\bs{\beta}^*_{PRD}$ with the  minimum overall unfitness.
\ei
\noin
\tb{Remarks 3.4}
\bi
\item[\tb{(I)}] Note that in (C) above, due to the objective function in (\ref{beta*.eqn}) is not differentiable w.r.t. $\bs{\beta}$, therefore many fancy gradient-type optimization methods are not applicable. However, downhill simplex method (Nelder-Mead), and other non-linear and non-convex optimization algorithms (such as MCMC and simulated annealing) could be used.
\vs
\item[\tb{(II)}] The algorithm above is essentially a modification of the one given in Z19, where it first searches for $(p+1)$ deepest sample points, then
over the convex hull formed by these $(p+1)$ points searches for the final $\bs{\beta}$. A drawback of the latter algorithm is the convex hull might be too small and misses the real deepest point $\bs{\beta}^*(F^n_{\mb{Z}})$.
\hfill \pend
\ei
\section{Examples}
Examples are presented below
for the illustration of the algorithms proposed in this article on the exact computation of PRD and approximate computation of its induced median $\bs{\beta}^*$.
\subsection{On the computation of PRD}
Here we want to compare the exact computation algorithms in Z19 and the one in this article. For the latter one, we now explain in detail the implementation of two types of optimization.\vs
Given a direction $\mb{v}\in P_k$, a permutation, say, $i_1,\cdots, i_n$ is obtained. That is,  for all the values from $\{k^{\mb{v}}_i=1/\mb{t}'_i\mb{v}\}$,
we have $k^{\mb{v}}_{i_1}\leq k^{\mb{v}}_{i_2}\leq \cdots \leq k^{\mb{v}}_{i_n}$. \tb{Type I} optimization problem could be described as
\bi
\item[]\tb{minimize}: $ 
\mb{c}'\mb{v}$, 
\item[]\tb{subject to}: (i) $\mb{B}'\mb{v}\leq \mb{0}_{(n-1)\times 1}$; ~~  (ii) $\mb{v}'\mb{v}=1$,
\ei
where $\mb{c}$ and $\mb{B}$ are constant vector and matrix, respectively (see Remarks 3.1 and (\ref{B.eqn})), $\min$ could also be $\max$. That is, we have a linear objective function, and a linear inequality constraint and a quadratic equality constraint.\vs
When $p=2$, each $P_k$ becomes a piece of arc of the unit circle and the cones formed by the linear constraints are the angular regions with two radii as their boundaries. The optimization problem becomes
linear programming over the piece of arc. By the fundamental theory of linear programming, the minimization
or maximization occurs only at the boundary. So only evaluation of $\mb{c}'\mb{v}$ is needed for $\mb{v}$ at the two boundary directions. There are at most $O(n^2)$ pieces of $P_k$'s. \vs Generally,
\tb{Type I} optimization problem can be solved by an augmented
Lagrangian minimization using R package `alabama’, or by sequential quadratic programming using
R solver `slsqp’. Alternatively, it can be 
transformed into semidefinite programming
problems and solved using R solver `csdp’. Also R packages ``optisolve" and ``nlopt" are applicable. \vs

Now we turn to the \tb{Type II} optimization problem. It could be describes as
\bi
\item[]\tb{minimize}: $ 
\frac{\mb{b}'\mb{v}}{\mb{v}'\mb{A}\mb{v}}$, 
\item[]\tb{subject to}: (i) $\mb{B}'\mb{v}\leq \mb{0}_{(n-1)\times 1}$; ~~  (ii) $\mb{v}'\mb{v}=1$,
\ei
where $\mb{b}$ and $\mb{A}_{p\times p}$, $\mb{B}$ are constant vector and matrices, respectively (see Remarks 3.1 and (\ref{B.eqn})), $\mb{A}$ could be treated as a symmetric and positive definite one, $\min$ could also be $\max$.\vs That is, we have a non-linear, non-convex, but differentiable objective function, or a rational objective function, and a linear inequality constraint and a quadratic equality constraint.
The problem again can be solved by using R packages `alabama’,  ``optisolve",  and ``nlopt".
\vs

\vs
In the following example, we examine the performance of exact (Z19 and Section \ref{EX-prd.section})
and approximate (AA-UF-3 of Z19) 
computation of UF, equivalently PRD, for a real data set.\vs 

\begin{figure}[h!]
\vspace*{-5mm}
\includegraphics
[width=\textwidth] {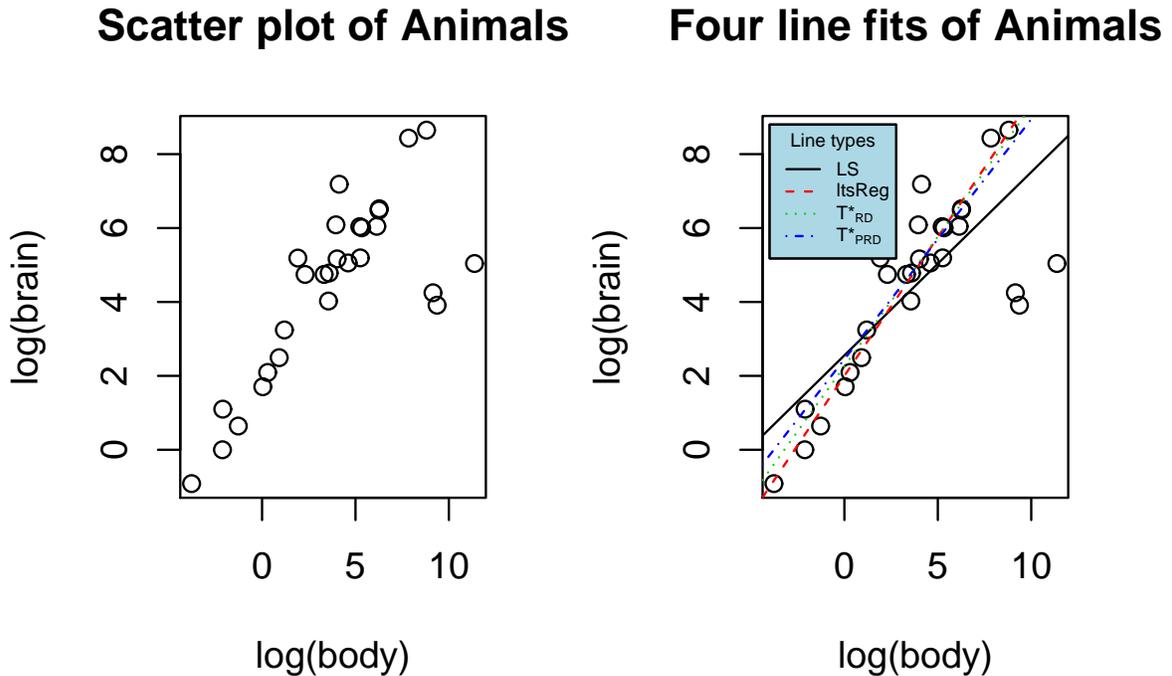}
\vspace*{-5mm}
 \caption{{Four regression lines based on the data of brain and body weight. Solid black for LS line; dashed red for ltsReg line, dotted green for $T^*_{RD}$; dot-dash blue for $T^*_{PRD}$.
  }}
 \label{fig:example-4-1}
 \vspace*{-8mm}
\end{figure}

\noin
\tb{Example 4.1.} Average of brain and body weight data (source: Table 7, page 58 of RL87).\vs
The average of brain weight (in grams) and the body weight (in kilograms) of 28 animals are investigated whether a larger
brain is required to govern a heavier body. A plot of original measurements is not very informative, a logarithmic transformation was necessary. The plot of the transformed data exhibits an overall linear relationship (see the left panel of Fig. \ref{fig:example-4-1}). It is clear that three outliers (dinosaurs) form the right lower cluster.
\vs
We regress the transformed data with four methods:
  LS (least squares); ltsReg (least trimmed squares (Rousseeuw (1984)); $T^*_{RD}$ (maximum regression depth (RD) (Rousseeuw and Hurbert (1999) (RH99) estimator, see section \ref{prd.section} for definition, see Rousseeuw and Struyf (1998) and  Liu and Zuo (2014) for computation); and $T^*_{PRD}$ (or $\bs{\beta}^*_{PRD}$) (see section \ref{prd.section} for computation).\vs The last two represent the maximum depth induced median type regression estimators whereas the first (LS) is the traditional one which is notorious for its non-robustness and the second one (ltsReg) represents the most robust and prevailing regression estimator.
\vs
Four lines (or four $\bs{\beta}$'s, $\bs{\beta}'=(\beta_0, \beta_1)=\mbox{(intercept, slope)}$) from the four methods are (2.55490, 0.49599), (2.00135, 0.75087), (2.258175, 0.7028644), and (2.45098, 0.64920), respectively. The first (LS) line is (slight different from the one given in RL87) apparently attracted by the outlier cluster downwards.
 Other three robust alternatives indeed resist to the outliers while the last two depth induced medians are almost identical (see the right panel of
 Fig. \ref{fig:example-4-1})
\bec
\enc
Note that there actually exit three deepest regression depth lines: (2.258175, 0.7028644);
(2.445328, 0.6677692) and (2.466361, 0.6501526), each possessing RD (of RH99): $12/28$.  The non-uniqueness issue of maximum regression depth estimator has been addressed in Zuo (2020).
\vs Note that the average of all three deepest RD lines is (2.38995, 0.67360). This is the line recommended in RH99. However, its regression depth  is $11/28$, no longer the maximum regression depth (or the line no longer fits ``the deepest regression method"). This phenomenon has been observed in  Mizera and Volauf (2002) and Van Aelst et al (2002).\vs

For the four $\bs{\beta}$'s, we calculate their UF's with the exact algorithms of Z19 (EA-Z19) and the one in Section
\ref{EX-prd.section} (denoted by EA-Z20) and approximate algorithm AA-UF-3 of Z19 (AA-Z19). The obtained UF and the consumed time are reported in the table below.\vs
UF induced rank (in ascending order) of each line is also reported. Regression depth (RD) of RH99 (see section \ref{prd.section}) of each line, as well as the induced rank (in descending order) are also reported. \vs

\vspace*{-5mm}
\bec
\begin{table}[h!]
\centering
Table entries (a,b,c) are a:= UF (or RD), b:=time consumed (in seconds), c:=induced rank.\\[2ex]
\begin{tabular}{c c c c c c } 
method   & LS & ltsReg & $T^*_{RD}$ & $T^*_{PRD}$\\[1ex]
\hline
UF(EA-Z19) &(1.365, 0.017, 4)&(0.637, 0.023, 3) &(0.407, 0.015, 2)&(0.347, 0.015, 1)\\[1ex]
UF(EA-Z20)& (1.286, 2.803, 4)& (0.569, 2.799, 3)  & (0.350, 2.779, 2) &(0.290, 2.776, 1)\\ [1ex]
UF(AA-Z19)&(1.285, 0.030, 4)& (0.569, 0.030, 3)& (0.332 ,0.030, 2)& (0.290, 0.031, 1)\\[1ex]
\hline
RD(RH99)&(4/28, 0.002, 4)&(10/28, 0.001, 3)& (12/28, 0.001, 1)& (11/28, 0.001, 2)\\[1ex]
\hline
\end{tabular}
\caption{~Performance of exact and approximate algorithms w.r.t. different $\bs{\beta}'s$ (lines). Four lines are ranked by different criteria}
\label{ex-aa.tab}
\end{table}
\vspace*{-10mm}
\enc
Table \ref{ex-aa.tab} consists of two parts. One part is about the unfitness, or equivalently, the projection depth and its induced rank and the consumed computation time of each method for four lines. The other part is about the same thing but based on regression depths for four lines which are obtained by utilizing the R package "mrfDepth" that utilizes R package Rcpp. 
\vs
\noin
\tb{Remarks 4.1} The table reveals that
\bi
\item[(I)]
Three methods EA-Z19, EA-Z20, and AA-Z19 yield the same induced rank of the four lines. Based on their UF, from the worst to the best, it is
LS, ltsReg, T$^*_{RD}$ and T$^*_{PRD}$.
\item[ (II)] EA-Z19 produces the largest UF in all four cases while the AA-Z19 yields UF's that very close to those of EA-Z20 (the results from AA-Z19 are very stable in the approximation for the different direction numbers used: $10^3$, $10^4$ or $10^5$. In the table it employed $10^3$) but always no greater than the latter. Generally speaking, the larger the UF obtained the more accurate the results are. This general principle indicates that EA-Z20 does its job
whereas EA-Z19 although it gives the largest UF's but they are not the most accurate.\vs How can that be?  The largest UF's are due to the modification of  the regular sample median in EA-Z19. The latter is modified to be the ``low median" in Z19. The low median is always less than the regular median with respect to projected values. However, its absolute value might be greater than that of the regular median if both the regular median and the low median of the projected values are negative in some direction. Consequently, they are the most inaccurate results. This indeed is the price EA-Z19 has to pay for its speed (see (III) below).\vs
\item[(III)] In terms of computation time for UF, EA-Z19 is surprisingly the fastest (and even faster than the AA-Z19), and EA-Z20 is the slowest. This is   due to the modification of the median in EA-Z19 which leads the optimization problem to the evaluation of UF along $O(n)$ directions (see Z19, the proof of Theorem 2.1).
\item[(IV)] In terms of regression depth ranking, LS and ltsReg are still the worst and the second-worst choices whereas the ranks of $T^*_{RD}$ and $T^*_{PRD}$ are switched, $T^*_{RD}$ becomes the only best choice as it is expected. This is no longer true if $T^*_{RD}$ is the average of the three deepest lines. (The comparisons here are somewhat unfair since if we look at the sum of residuals squares, then LS becomes the best choice. Likewise, ltsReg could also become the best if the comparison criterion is changed.) \hfill \pend \vs
\ei
All results above (and below) were obtained on a desktop Intel(R)Core(TM) i7-2600 CPU @ 3.40GHz 3.40GHz. AA-UF-Z19 employed matlab code.
R code and matlab code are downloadable via https://www.stt.msu.edu/users/zuo/Codes/2020/readme-Z20.txt.

\vs
\subsection{On the computation of PRD induced median}\label{prd.section}
The most famous notion of depth in regression and its induced median are regression depth of Rousseeuw and Hurbert (1999) (RH99) and its induced median, respectively.
\vs
For any $\bs{\beta}\in \R^p$ and the joint distribution $P$ of $(\mb{x}', y)'$ in (\ref{eqn.model}), RH99 defined the regression depth of $\bs{\beta}$, denoted hence by $\mbox{RD}_{RH} (\bs{\beta};P)$,
to be the minimum probability mass that needs to be passed when tilting (the hyperplane induced from) $\bs{\beta}$ in any way until it is vertical. The maximum regression depth functional $\bs{\beta}^*_{{RD}_{RH}}$ (also denoted by $T^*_{RD}$ or $\bs{\beta}^*_{RD}$) (aka regression median) is defined as
\be \bs{\beta}^*_{{RD}_{RH}}(P)=\arg\!\max_{\bs{\beta}\in\R^p}\mbox{RD}_{RH}(\bs{\beta};P) \label{T-RD.eqn}
\ee

Many characterizations of $\mbox{RD}_{RH}(\bs{\beta};P)$, or equivalent definitions, have been given in the literature, see, e.g., Z18 and references cited therein.\vs

 \vs

\vspace*{-0mm}
\begin{table}[h!]
\centering
~~ Table entries: (empirical mean squared error, average time per sample (seconds))
\bec
\begin{tabular}{c c c c c c}
n& method & $p=2$~~~~ & $p=3$~~~~ &$p=4$~~~~& $p=6$~~~~  \\
\hline\\[0.ex]
$40$& $\bs{\beta}^*_{PRD} (Z19)$ &(0.244, 7.424)&(0.488, 18.69)& (0.737, 13.21) & (1.505, 12.01)\\[.5ex]
 & $\bs{\beta}^*_{PRD} (Z20)$& (0.232, 0.060) &(0.468, 0.261)&(0.723, 0.304)&(1.429, 0.354)\\[.5ex]
 &$\bs{\beta}^*_{RD}$ &(0.243, 0.038)&(0.492, 0.124)& (2.7e+04, 6.542)& (1.717, 9.619)\\[.5ex]
 &ltsReg &(0.380, 0.007)&(0.579,  0.011)& (0.781, 0.010)& (1.434, 0.018)\\[.5ex]\\
$60$& $\bs{\beta}^*_{PRD}(Z19)$ &(0.172, 9.076)&(0.339, 22.04)& (0.543, 19.70)& (0.986, 22.82)\\[.5ex]
 & $\bs{\beta}^*_{PRD} (Z20)$& (0.160, 0.080) &(0.323, 0.310)&(0.510, 0.445)&(0.894, 0.532)\\[.5ex]
  &$\bs{\beta}^*_{RD}$ &0.172, 0.043)&(0.366, 0.286)& (2565.1, 23.14)& (1.206, 11.82)\\[.5ex]
  &ltsReg &(0.326, 0.007)&(0.475, 0.013)& (0.599, 0.015)& (0.894, 0.024)\\[.5ex]\\
$80$& $\bs{\beta}^*_{PRD} (Z19) $& (0.131, 10.29) &(0.273, 26.82)&(0.428, 25.00)&(0.821, 26.17)\\[.5ex]
  & $\bs{\beta}^*_{PRD} (Z20)$& (0.124, 0.100) &(0.260, 0.436)&(0.413, 0.613)&(0.691, 0.634)\\[.5ex]
 &$\bs{\beta}^*_{RD}$&(0.130, 0.047) &(0.291, 0.569)&(2012.6, 58.42)& (1.111, 14.08)\\[.5ex]
 &ltsReg&(0.290, 0.009) &(0.416, 0.018)&(0.506, 0.020)& (0.703, 0.029)\\[.5ex]\\
$100$& $\bs{\beta}^*_{PRD}(Z19)$ &(0.108, 10.22)&(0.233, 28.90)& (0.370, 28.63)& (0.655, 31.40)\\[.5ex]
 & $\bs{\beta}^*_{PRD} (Z20)$& (0.100, 0.123) &(0.221, 0.528)&(0.346, 0.687)&(0.555, 0.763)\\[.5ex]
  &$\bs{\beta}^*_{RD}$ &(0.109, 0.048)&(0.252, 0.950)& (5.5e+06, 101.8)& (0.963, 16.37)\\[.5ex]
  &ltsReg &(0.252, 0.010)&(0.418, 0.021)& (0.455, 0.024)& (0.578, 0.035)\\[1ex]
\hline
\end{tabular}
\enc
\caption{Performance of different regression methods for various $n$ and $p$.}
\label{table-comp-time}
\end{table}
\vspace*{-5mm}
As a median in regression,  $\bs{\beta}^*_{{RD}_{RH}}(P)$ is a promising robust alternative to the classic least squares (LS) regression estimator. In fact,
in terms of asymptotic breakdown point (ABP) robustness, the former possesses a $33\%$ ABP (Van Aelst and Rousseeuw (2000) (VAR00)), in contrast to $0\%$ of the latter. \vs
Zuo (2019b) (Z19b) has investigated the ABP of $\bs{\beta}^*_{PRD}$, it turns out that it possesses the highest possible ABP, $50\%$. For this advantage over
 $\bs{\beta}^*_{{RD}_{RH}}$ (see illustration examples in Z19b), it has to pay a price in the computation. The cost of the computation of  $\bs{\beta}^*_{{RD}_{RH}}$ is generally lower than that of $\bs{\beta}^*_{PRD}$.
 \vs
 To see the difference in the computation cost, we list below the computation time consumed by both medians for different sample sizes n and dimensions d. For the benchmark and comparison purpose, we also list the times consumed by the famous least trimmed squares
(Rousseeuw (1984)) regression (ltsReg) estimator and the times consumed by $\bs{\beta}^*_{PRD}$ in Z19 (denoted by $\bs{\beta}^*_{PRD}(Z19))$.
The latter is the one of deepest hyperplanes obtained by searching the convex hull formed
by $(p+1)$ deepest candidate $\bs{\beta}’$s (see Section 5). Function rdepth in R package ”mtfDepth” was used to calculate the RD of each candidate hyperplane. The performance of four algorithms for $\bs{\beta}^*_{RD_{RH}}$, $\bs{\beta}^*_{PRD}(Z19)$, $\bs{\beta}^*_{PRD}$  in Section 3.2 (denoted by $\bs{\beta}^*_{PRD}(Z20)$)), and ltsReg,
respectively, is demonstrated in the table \ref{table-comp-time}.  \vs

We generate $1000$ samples $\mb{Z}^{(n)}=\{(\mb{x}'_i, y_i), i=1,\cdots, n, \mb{x}_i \in \R^{p-1}\}$ from the Gaussian distribution with zero mean vector and $1$ to $p$ as its diagonal entries of the diagonal covariance matrix  for various $n$ and $p$. They are contaminated by $5\%$ i.i.d. normal $p$-dimensional points with individual mean $10$ and variance $0.1$. Thus, we no longer have a symmetric errors and homoscedastic variance model
(skewness and heteroscedasticity are allowed for RD of RH99).\vs

For a general estimator $\mb{T}$, if it 
 is regression equivariant, then we can assume (w.l.o.g.) that the true parameter $\bs{\beta}_0=\mb{0}\in \R^p$.  We calculate
$\mbox{EMSE}:=\frac{1}{R}\sum_{i=1}^R \|\mb{T}_i - \bs{\beta}_0\|^2$, the empirical mean squared error (EMSE) for $\mb{T}$, where
$R = 1000$, $\bs{\beta}_0 = (0, \cdots, 0)'\in \R^{p}$,
 and $\mb{T}_i$ is the realization of $\mb{T}$ obtained from the ith sample with size $n$. The EMSE and the average computation time (in seconds) per sample by different estimators are listed in Table \ref{table-comp-time}.\vs

\noindent
\tb{Remarks 4.2} Table \ref{table-comp-time} reveals that
\bi
\item[(I)] In terms of the average time consumed per sample, or computation speed, (i) the ltsReg  is the fastest in all cases whereas the $\bs{\beta}^*_{RD}$ is the second fast method when $p$ is $2$, or $3$ (and $n\leq 60$).
  (ii)  $\bs{\beta}^*_{PRD}$(Z19) is the slowest in almost all cases with exceptions in $p=4~(n>40)$  cases where $\bs{\beta}^*_{RD}$ unexpectedly becomes the slowest. (iii) $\bs{\beta}^*_{PRD}$(Z20) is at least 30 times faster than  $\bs{\beta}^*_{PRD}$(Z19) in all cases, sometimes (p=2) it is more than 100 times faster. It is also at least 20 times faster than $\bs{\beta}^*_{RD}$ when $p>3$.\vs 
  Note the comparison here is somewhat unfair to $\bs{\beta}^*_{PRD}$(Z19) since it is the only one that utilizes purely R programming for the entire calculation whereas ltsReg using Fortran and
     $\bs{\beta}^*_{RD}$ and $\bs{\beta}^*_{PRD}$(Z20) employing Rcpp in the background computation. This example also confirms that old Forthan is still an
     excellent programming language for scientific computation.

\item[(II)] Computation speed is just one of the important performance criteria. Accuracy or efficiency is another, if not more important one.
In terms of EMSE, there is  an across-board winner. That is, $\bs{\beta}^*_{PRD}$(Z20) has the smallest EMSE in all cases
considered.
\item[(III)] In terms of speed and EMSE, $\bs{\beta}^*_{PRD}$(Z20) outperforms $\bs{\beta}^*_{PRD}$(Z19) in all cases. Furthermore, the former consumes less than one second in all cases considered. 
     \hfill \pend
\ei
\vs  The ltsReg has a fairly good finite sample relative efficiency, but it is also notorious
    for its  inefficient in the asymptotic sense (with asymptotic efficiency just $7\%$ (see Stromberg, et al.(2000)). It benefits from Fortran for its speed.
In the sequel, ltsReg will be excluded from our discussion for a pure apple vs apple (depth median vs depth median) fair comparison.
\vs

\begin{table}[b!]
\centering
Replication $1000$ times,  $n=65$  \\[1ex]
\bec
\begin{tabular}{c c c c c }
 Performance criteria~~~ &$\bs{\beta}^*_{PRD} (Z19)$~~~ & $\bs{\beta}^*_{PRD} (Z20)$~~~~&$\bs{\beta}^*_{RD}$ 
  \\[.5ex]
\hline\\[0.ex]
&{\bf{Case I}}& $p=3$ &  &\\[1.5ex]
EMSE&0.10434764 &0.09433006 &0.11191986 
\\[.5ex]
Time consumed per sample &21.14003496 &0.36948846&0.34871839 
\\[.5ex]
\hline\\[.ex]
&{\bf{Case II}}& $p=4$ & &\\[1.5ex]
EMSE&0.1652269&0.1516346&5657894 
\\[.5ex]
Time consumed per sample &12.75727514 &0.26809841 &26.00944714 
\\[.5ex]
\hline\\[.ex]
&{\bf{Case III}}& $p=5$ & &\\[1.5ex]
EMSE&0.2622625 &0.2372195 &0.2519083
\\[1.5ex]
Time consumed per sample &13.41192399 &0.22595816 &6.72852676 
\\[.5ex]
\hline\\[0.ex]
\end{tabular}
\enc
\vspace*{-9mm}
\caption{Performance of different regression depth medians for three true $\bs{\beta}_0$'s.}
\label{table-3-betas}
\vspace*{-0mm}
\end{table}

\noin
\tb{Example 4.2} Now we investigate the performance of the three regression depth medians ($\bs{\beta}^*_{PRD}(Z19)$, $\bs{\beta}^*_{PRD}(Z20)$, and $\bs{\beta}^*_{RD}$) in a slightly different setting. We generate $1000$ samples $\{(\mb{x}'_i, y_i) \in \R^p\}$ with a fixed sample size $65$ from an assumed model: $y=\bs{\beta_0}'\mb{x}     
+e$, where $\mb{x}=(1,x_1,\cdots, x_{p-1})'$ and $\bs{\beta_0}=(\beta_0,\cdots, \beta_{p-1})'$ are in $\R^p$ and $x_i$ and e are from either Cauchy or standard Gaussian distribution.\vs  We list the average time consumed (in seconds) per sample and the EMSE (the same formula as before) for the three methods
with respect to different $\bs{\beta_0}$'s in Table \ref{table-3-betas}.  {\bf{Case I}} $\bs{\beta_0}=(-2, 0.1,1)'$, all $x_i$ and $e$ are from $N(0,1)$ distribution.
{\bf{Case II}} $\bs{\beta_0}=(-2, 0.1,1, 5)'$, $x_1$ is from $N(0,1)$ and all other $x_i$ and $e$ are from Cauchy distribution. {\bf{Case III}} $\bs{\beta_0}=(50, 0.1, -2, 15, 100)'$, all $x_i$ and $e$ are from $N(0,1)$ distribution.
\vs

Inspecting table 3 reveals that (i) $\bs{\beta}^*_{PRD}(Z20)$ is much (ranging from $47-59$ times) faster than the slower $\bs{\beta}^*_{PRD}(Z19)$ in all cases, it is also $97$ and $29.78$  times faster than $\bs{\beta}^*_{RD}$ in the cases of $p=4$ and $p=5$, respectively, (ii) $\bs{\beta}^*_{PRD}(Z20)$ has the smallest EMSE as well in all cases, (iii) the sample variance (or more precisely EMSE) of both PRD induced medians increases when $p$ increase whereas the time consumed per sample for the fixed sample size by $\bs{\beta}^*_{PRD}(Z20)$ decreases in this case.
\vs
All results above and below are obtained on a desktop Intel(R)Core(TM) i7-2600 CPU @ 3.40GHz 3.40GHz. 
To download R codes in this and the next sections, utilizing the link: https://www.stt.msu.edu/users/zuo/Codes/2020/readme-Z20.txt.\vs
\section{Other estimators induced from PRD}
Before introducing other estimators, we like to first explain why $\bs{\beta}^*_{PRD}(Z20)$ runs faster than  $\bs{\beta}^*_{PRD}(Z19)$.
First, we briefly review the main computation steps of  $\bs{\beta}^*_{PRD}(Z19)$ (cf. Section 3.2 (A)-(D))
\bi
\item[(i)] Generating $N_{\bs{\beta}}$ $\bs{\beta}$'s via the hyperplane $y=\mb{x}'\bs{\beta}$ based on $p$ points sampled from $\mb{Z}^{(n)}:=\{(\mb{x}'_i, y_i), i=1,\cdots, n\}$,
where $N_{\bs{\beta}}$ is a tuning parameter and never greater than $n\choose p$.
\item[(ii)] Computing the unfitness (UF) for each $\bs{\beta}$ using special directions (including those perpendicular to $\mb{t}_i=\mb{x}_i/r_i(\bs{\beta})$ where $r_i(\bs{\beta})=y_i-\mb{x}'_i\bs{\beta}$, and those $p$ axis directions, and those $N_{\mb{v}}$ normal directions of hyperplane formed as those in (i) by $p$ sample points from $\mb{Z}^{(n)}$, where $N_{\mb{v}}$ is another tuning parameter which increases when $p$ increases.
\item[(iii)]   After the computation of UF for $(p+1)$ $\bs{\beta}$'s in step (ii) above, calculating the minimum UF (UF-min), and updating this UF-min after each computation of UF of candidate $\bs{\beta}$ and using it to skip the computation of some candidate $\bs{\beta}$'s if along some direction, the one-dimensional unfitness of the $\bs{\beta}$ (see the RHS of (\ref{UF.eqn}) or (13) of Z18) is greater than the UF-min since this $\bs{\beta}$ can never be final solution which shall have a global minimum UF. This UF-min cuts a tremendous amount of unnecessary computation cost.
\item[(iv)] Selecting Nbet (another tuning parameter) $\bs{\beta}$'s from
the convex hull formed by $(p+1)$ deepest (or equivalently with minimum UF) $\bs{\beta}$'s. The deepest $\bs{\beta}$ among the Nbet $\bs{\beta}$ is
treated as the final $\bs{\beta}^*_{PRD}(Z19)$.
\ei
\vs
$\bs{\beta}^*_{PRD}(Z20)$  has almost the same steps but with different details. For example, in (iii) above, Z19 computes UF for $N$ ($\leq N_{\bs{\beta}}$) $\bs{\beta}$'s, each time it samples a $\bs{\beta}$ from the candidate $\bs{\beta}$ matrix B ($N_{\bs{\beta}}$ by $p$) constructed from (i), and after finishing
$(p+1)$ computations, it calculates the minimum unfitness (UF-min) of all $(p+1)$ UF's, then updates the UF-min after each computation of UF utilizing a nested if and else statement.
\vs
$\bs{\beta}^*_{PRD}(Z20)$ first skips the sampling step and just directly invokes the $\bs{\beta}$ from matrix B, and it replaces the nested if and else statement by a simple if statement. $\bs{\beta}^*_{PRD}(Z20)$ also uses min function to replace sort function in the search over the convex hull for the final solution. These simple steps boost the computation speed five times. Furthermore, $\bs{\beta}^*_{PRD}(Z20)$ employs Rccp package 
which eventually resulted in its speed is at least 30 times faster than $\bs{\beta}^*(Z19)$.\vs

Computational speed is not the only improvement of $\bs{\beta}^*_{PRD}(Z20)$, it also cuts the EMSE of $\bs{\beta}^*(Z19)$. To achieve this goal, $\bs{\beta}^*_{PRD}(Z20)$ takes the advantage of the solution from ltsReg and the deepest  $\bs{\beta}$'s  with maximum RD (might not be a unique one, but all are also from $B$ which is shared by $\bs{\beta}^*_{RD}$) and adds them (a sub-matrix $B_1$) to the B matrix. It not only searches over the convex hull formed by $(p+1)$ deepest $\bs{\beta}$'s with minimum UF from $B$ but also considers the combinations of member of $B_1$. The final $\bs{\beta}$ with minimum UF is the solution of $\bs{\beta}^*_{PRD}(Z20)$. For more details, see the code posted on the link mentioned before.
\vs
$\bs{\beta}^*_{PRD}(Z20)$ is much faster than $\bs{\beta}^*(Z19)$, are there any depth induced estimators that run even faster than $\bs{\beta}^*_{PRD}(Z20)$?
From the discussion above, there are obviously other projection regression depth (PRD) induced estimators that can be computed even faster.
\vs

 The first one adds no extra computation cost to already obtained candidate $\bs{\beta}$ matrix $B$, it is just the deepest $\bs{\beta}$ with minimum UF in matrix $B$, denoted it by $\bs{\beta}^*_{PRD1}$. The second one is the plain average of deepest $(p+1)$ $\bs{\beta}$'s from $B$, denoted it by $\bs{\beta}^*_{PRD2}$.
 The third one is a UF weighted estimator defined below, denoted it by $\bs{\beta}^*_{PRD3}$,
\be
\bs{\beta}^*_{PRD3}=\frac{\sum_{i=1}^{(p+1)}w(\rho_i)\bs{\beta}_{(i)}}{\sum_{i=1}^{(p+1)}w(\rho_i)}, \label{wprd.eqn}
\ee
where $\rho_i=\mbox{UF}(\bs{\beta}_{(i)})$ and $\bs{\beta}_{(1)},\cdots,\bs{\beta}_{(p+1)}$ are first $(p+1)$ deepest $\bs{\beta}$'s (with least UF) in B and the weight function $w$ is defined as follows:
\be
w(r)={\bf{I}}(r\leq r_0)+{\bf{I}}(r>r_0)\frac{exp~\Big(k\big(2r_0/r-(r_0/r)^2\big)\Big)-1}{exp~ (k)-1}, \label{weight.eqn}
\ee
with two turning parameters $k$ and $r_0$, we set $k=3$ and $r_0=\rho_{(p-1)}$, the $(p-1)$th smallest UF among the $(p+1)$ minimum UF's.
For more discussion on this weight function and the tuning parameters, refer to Zuo (2003) and Z19b.
\vs
These estimators obviously can run faster than $\bs{\beta}^*_{PRD}(Z20)$ since they skip the time-consuming step of searching over the convex hull. One naturally wonders what are their EMSE's?\vs
Next, we investigate the performance of $\bs{\beta}^*_{PRD}(Z19)$, $\bs{\beta}^*_{PRD}(Z20)$, $\bs{\beta}^*_{PRD1}$, $\bs{\beta}^*_{PRD2}$, and $\bs{\beta}^*_{PRD3}$. For the benchmark purpose, the famous depth median: $\bs{\beta}^*_{RD}$ of RH99 is included in the comparison. $1000$ samples are generated with the same scheme as that for table 2.
 \vs
\vspace*{-0mm}
\begin{table}[h!]
\centering
~~ Table entries: (empirical mean squared error, average time per sample (seconds))
\bec
\begin{tabular}{c c c c c c}
n& method & $p=2$~~~~ & $p=3$~~~~ &$p=4$~~~~& $p=6$~~~~  \\
\hline\\[0.ex]
$40$& $\bs{\beta}^*_{PRD}(Z19) $ &(0.249, 7.289)&(0.465, 9.083)& (0.743, 8.144) & (1.493, 12.06)\\[.5ex]
 & $\bs{\beta}^*_{PRD}(Z20) $& (0.237, 0.062) &(0.448, 0.142)&(0.736, 0.208)&(1.373, 0.343)\\[.5ex]
 & $\bs{\beta}^*_{PRD1} $& (0.244, 0.023) &(0.481, 0.040)&(0.831, 0.068)&(1.646, 0.142)\\[.5ex]
 & $\bs{\beta}^*_{PRD2} $& (0.268, 0.023) &(0.489, 0.040)&(0.882, 0.068)&(1.431, 0.142)\\[.5ex]
 & $\bs{\beta}^*_{PRD3} $& (0.258, 0.023) &(0.476, 0.040)&(0.771, 0.068)&(1.375, 0.142)\\[.5ex]
 &$\bs{\beta}^*_{RD}$ &(0.240, 0.040)&(0.466, 0.124)& (3195.3, 6.507)& (1.678, 9.382)\\[.5ex]\\
$60$& $\bs{\beta}^*_{PRD}(Z19)$ &(0.164, 9.140)&(0.346, 11.19)& (0.552, 9.564)& (1.050, 7.139)\\[.5ex]
 & $\bs{\beta}^*_{PRD}(Z20) $& (0.157, 0.082) &(0.329, 0.187)&(0.519, 0.268)&(0.923, 0.193)\\[.5ex]
 & $\bs{\beta}^*_{PRD1} $& (0.167, 0.031) &(0.363, 0.051)&(0.613, 0.090)&(1.139, 0.088)\\[.5ex]
 & $\bs{\beta}^*_{PRD2} $& (0.188, 0.031) &(0.484, 0.051)&(0.603, 0.090)&(1.131, 0.088)\\[.5ex]
 & $\bs{\beta}^*_{PRD3} $& (0.175, 0.031) &(0.446, 0.051)&(0.568, 0.090)&(1.075, 0.088)\\[.5ex]
  &$\bs{\beta}^*_{RD}$ &(0.165, 0.043)&(0.350, 0.300)& (4703.0, 21.18)& (1.337, 8.585)\\[.5ex]\\
$80$& $\bs{\beta}^*_{PRD}(Z19)  $& (0.135, 9.371) &(0.284, 27.79)&(0.446, 25.66)&(0.795, 9.229)\\[.5ex]
  & $\bs{\beta}^*_{PRD}(Z20) $& (0.128, 0.101) &(0.261, 0.441)&(0.412, 0.611)&(0.666, 0.288)\\[.5ex]
  & $\bs{\beta}^*_{PRD1} $& (0.134, 0.040) &(0.297, 0.095)&(0.492, 0.165)&(0.832, 0.129)\\[.5ex]
  & $\bs{\beta}^*_{PRD2} $& (0.165, 0.040) &(0.315, 0.095)&(0.509, 0.165)&(0.872, 0.129)\\[.5ex]
   & $\bs{\beta}^*_{PRD3} $& (0.147, 0.040) &(0.302, 0.095)&(0.481, 0.165)&(0.830, 0.129)\\[.5ex]
 &$\bs{\beta}^*_{RD}$&(0.132, 0.047) &(0.291, 0.583)&(4446.2, 58.50)& (1.050, 10.64)\\[.5ex]\\
$100$& $\bs{\beta}^*_{PRD}(Z19)$ &(0.121, 10.24)&(0.237, 14.73)& (0.387, 27.63)& (0.698, 11.79)\\[.5ex]
 & $\bs{\beta}^*_{PRD}(Z20)$& (0.109, 0.121) &(0.218, 0.301)&(0.361, 0.719)&(0.551, 0.338)\\[.5ex]
 & $\bs{\beta}^*_{PRD1} $& (0.117, 0.048) &(0.247, 0.086)&(0.439, 0.202)&(0.682, 0.148)\\[.5ex]
 & $\bs{\beta}^*_{PRD2}$& (0.153, 0.048) &(0.275, 0.086)&(0.467, 0.202)&(0.851, 0.148)\\[.5ex]
 & $\bs{\beta}^*_{PRD3}$& (0.142, 0.048) &(0.263, 0.086)&(0.437, 0.202)&(0.771, 0.148)\\[.5ex]
  &$\bs{\beta}^*_{RD}$ &(0.115, 0.050)&(0.240, 0.960)& (2427164, 113.4)& (0.970, 12.24)\\[.5ex]
\hline
\end{tabular}
\enc
\caption{Performance of regression depth induced estimators for various $n$ and $p$.}
\label{table-comp-time-2}
\end{table}
\vspace*{-3mm}

\vs
Inspecting the table \ref{table-comp-time-2} immediately reveals that (i) $\bs{\beta}^*_{PRD}(Z20)$  has the smallest EMSE in  all cases
and it is at least $34$ (sometimes more than $100$) times faster  than $\bs{\beta}^*_{PRD}(Z19)$;   (ii) $\bs{\beta}^*_{PRD}(Z19)$ is the slowest (with the exceptions in $p=4$, $p=6$ and $n>40$ cases where $\bs{\beta}^*_{RD}$ becomes the slowest).  (iii) $\bs{\beta}^*_{PRD1}$, $\bs{\beta}^*_{PRD2}$, and $\bs{\beta}^*_{PRD3} $ are the fastest ($86$ to $300$ times faster than $\bs{\beta}^*_{PRD}(Z19)$)  and they are currently regarded as having the same speed (all depend on the given matrix $B$ of candidate $\bs{\beta}$'s and their unfitness and then on the sorted values of their unfitness).
 Among the three, the deepest of all $\bs{\beta}$ in $B$, $\bs{\beta}^*_{PRD1}$, and the depth weighted deepest $(p+1)$ $\bs{\beta}$'s, $\bs{\beta}^*_{PRD3}$  seemingly perform better and the plain average of them, $\bs{\beta}^*_{PRD2}$, seemingly performs worst in most cases.
Furthermore, our empirical evidence indicates that $\bs{\beta}^*_{PRD3}$ performs even better when $p$ increases (say $p\geq 8$).   (iv) Overall, $\bs{\beta}^*_{PRD}(Z20)$ should be recommended among the six depth induced regression estimators, it becomes empirically the same as $\bs{\beta}^*_{PRD1}$ for large $p$ (e.g. $p=20$, $n=40, 60, 80$), the second one should be recommended is the
$\bs{\beta}^*_{PRD3}$ (or $\bs{\beta}^*_{PRD}(Z19)$), and  $\bs{\beta}^*_{PRD2}$ could be abandoned.
\vs
\section{Concluding comments}
Unlike Z19, this article presents the exact  algorithm for the computation of the UF (or equivalently the PRD)
without modifying the original definition of univariate median and thus without scarifying invariance of projection regression depth and the equivariance of the depth induced median.  The second major contribution is to boost the speed of computation of the $\bs{\beta}^*_{PRD}(Z19)$ by at least $30$ times,
more importantly to reduce the empirical mean squared error of the depth induced regression median meanwhile.\vs
The article also introduces three regression depth induced estimators that can run even faster,  $86$ to $300$ times faster than $\bs{\beta}^*_{PRD}(Z19)$. These estimators satisfy regression, scale, and affine equivariance (see Z18 for definitions) and more importantly have roughly the same level of empirical mean squared errors as that of the latter.  \vs
The major motivation of introducing depth induced regression estimators is to provide  robust alternatives to the traditional least squares estimator  and to overcome the  non-robustness fatal drawback of the latter.
The three depth induced regression estimators are expected to be highly robust, just like the $\bs{\beta}^*_{PRD}$ in Z19b with high finite sample breakdown point. Detailed investigation of robustness and other properties of the three deserves to be pursued independently and elsewhere through.\vs 
Finally, in light of five PRD induced estimators in Table 4, one can even introduce another estimator which is the one among the three of the five (or all five) with minimum unfitness. Call this estimator as $\bs{\beta}^*_{PRD4}$. Its performance and properties is worthy of a thorough examination.
 \vs
\begin{center}
{\textbf{\large Acknowledgments}}
\end{center}
The author thanks Hanshi Zuo, Yan-Han Chen, and Dr. Wei Shao for their proofreading of the manuscript and useful discussions on C++, R, and  Rcpp programming, all of which have led to improvements in the manuscript.

\vs

\end{document}